\documentstyle[aps,prl,epsfig]{revtex}

\begin{document}

\wideabs{

\title{Nucleation of bended vortices in Bose-Einstein condensates in elongated traps.}
  
\author{Juan J. Garc\'{\i}a-Ripoll and V\'{\i}ctor M. P\'erez-Garc\'{\i}a}
  
\address{Departamento de Matem\'aticas, Escuela T\'ecnica Superior de
  Ingenieros Industriales\\ Universidad de Castilla-La Mancha, 13071 Ciudad
  Real, Spain }

\date{\today}
  
\maketitle
  
\draft


\begin{abstract}
  We study the generation of vortices in rotating axially elongated
  magneto-optical traps, a situation which has been realized in a recent
  experiment (K. W. Madison, F.  Chevy, W.  Wohlleben, J. Dalibard, Phys. Rev.
  Lett. {\bf 84} 806 (2000)). We predict that at a critical frequency the
  condensate experiences a symmetry breaking and changes from a convex
  cloud to a state with one bended vortex. We also discuss several effects which
  enlarge the critical frequency with respect to other geometries of the trap:
  these are, (i) the failure of the Thomas-Fermi approximation on the
  transverse degrees of freedom of the condensate, (ii) the enhancement of the
  transverse asymmetry of the trap by means of rotation and (iii) the yet
  unobserved bending of the vortex lines.
\end{abstract}

 }


Since the creation of the first Bose-Einstein condensates (BEC) using ultracold
dilute gases \cite{Siempre} and the success in describing them using simple but
accurate mean field theories, it has been a goal of the field to produce and
understand the Physics behind vortices in these condensates.

Vortices, or vortex-lines we should rather say, constitute the most relevant
topological defect that one finds in Physics. They basically consist on a twist
of the phase of a wave function around an open line and they are typically
associated to a rotation of a fluid whatever the fluid is made of (real
fluids, optical fluids, quantum fluids, ...)  \cite{vortices_real}.
Furthermore, vortices are one of the means by which quantum systems may acquire
angular momentum and thus react to some perturbations of the environment. They
have already been predicted, observed and studied in the superfluid phase of
$^4$He and are indeed known to be the key to some important processes inside
these systems, such as dissipation, moments of inertia and breakdown of
superfluidity. This is why extensive research on vortex generation, stability
and dynamics has been conducted in the field of BEC in the last years
\cite{vortices-BEC,Rokshar,us-vortex,PRL2000}.

In a recent experiment performed by a group from the \'Ecole Normale
Sup\'erieure (ENS) \cite{ENS} vortices are created by rotating an elongated trap
slightly deformed in the trasverse dimensions. This rotation is
performed at a constant angular speed, $\Omega$, which is the key parameter for
the appearance of one or more vortices. Many interesting phenomena have been
observed in these experiments, but we are concerned with some of them which are
controversial: (i) the first vortex nucleates at a critical rotation speed or
\emph{critical frequency}, $\Omega_1$, larger than the estimate of
$\Omega_1^{TF}=\frac{5\hbar}{2mR_\perp^2}\ln\frac{0.671 R_\perp}\xi$
\cite{Lundh}, (ii) vortex cores seem partially filled and (iii) after the
nucleation of the first vortex the angular momentum seems to depend
continuously on the rotation speed.

\begin{figure}
\begin{center}
  \epsfig{file=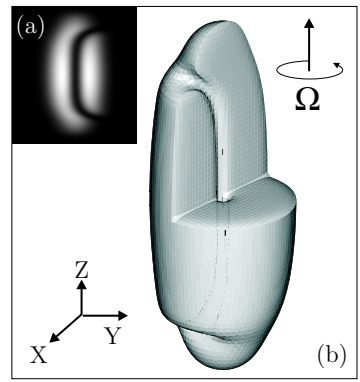,width=0.55\linewidth}
\end{center}
\caption{\label{agujero}
  Shape of the ground state of the Hamiltonian (\ref{energy}) for
  $\Omega = 0.75 \omega_{\perp}, \gamma = 350$, $Ng = 1000$ and $\varepsilon=0$.
  (a) Trasverse section. (b) Density plot. One quarter of the condensate has been removed to allow the direct
  observation of the inner structure of the vortex line.}
\end{figure}

In this letter we consider these effects within the framework of a mean field
theory for dilute gaseous condensates in rotating traps. In addition to an
explanation of the experimental features we find the surprising fact that above
$\Omega_1$ a symmetry breaking changes the ground state (GS) into a {\em bended} vortex
[Fig. \ref{agujero}].  We will discuss this phenomenon later.

\emph{The model.-} For most of the current experiments it is an accurate
approximation to use a zero temperature many-body theory of the condensate. In
that limit the whole condensate is described by a single wavefunction
$\psi({\bf r},t)$ ruled out by a Gross-Pitaevskii equation (GPE).

In the ENS experiment the trap is initially harmonic with axial symmetry, but
then a laser is applied which deforms it and makes it rotate with uniform
angular speed $\Omega$. By using the mobile reference frame which rotates with
the trap, we can describe this experiment using a modified GPE which reads
\begin{equation}
\label{GPE-rot}
i\frac{\partial \psi}{\partial t}  =\left[ -\frac{1}{2}\triangle +
V_{0}({\bf r}) +g\left| \psi \right| ^{2}
-\Omega L_{z}\right] \psi.
\end{equation}
Here the ENS trapping potential is given by
\begin{equation}
V_0({\bf r}) =
\frac{1}{2}\omega_\perp^2(1-\varepsilon)x^2+
\frac{1}{2}\omega_\perp^2(1+\varepsilon)y^2+
\frac{1}{2}\omega_z^2z^2.
\end{equation}
and $L_{z}=i\left( x\partial_{y}-y\partial _{x}\right)$ is a linear operator
which represents the angular momentum along the z-axis.

In Eq. (\ref{GPE-rot}) we have already applied a convenient adimensionalization
which uses the harmonic oscillator length,
$a_\perp=\sqrt{\hbar/m_{Rb}\omega_\perp}$ and period,
$\tau =\omega_\perp^{-1}.$ Using these units the nonlinear parameter becomes
$g=a_{\text{S}}/a_\perp$, where $a_{\text{S}}\simeq 5.5\, \textrm{nm}$ is the
scattering length for $^{87}$Rb, which is the gas used in Ref. \cite{ENS}.
Following the experiment we will take $\omega_z=2\pi \times 11.6\, \textrm{Hz}$
and $\omega_\perp=2\pi \times 232\, \textrm{Hz}$. Through the paper we will
also use a fixed value of $Ng = 1000$ which corresponds to a few times $10^5$
atoms.  For the small transverse deformation of the trap we have used
$\varepsilon = 0.03,0.06,0.09$ ($\varepsilon=0.03$ is the closest one to the
actual experiment).

The norm, $N[\psi]=\int \left| \psi \right| ^{2}d{\bf r},$ which can be
interpreted as the number of bosons in the condensate is a conserved quantity
as well as the energy
\begin{equation}
\label{energy}
E[\psi ]=\int \bar{\psi }\left[ -\frac{1}{2}\triangle +V_{0}\left( \mathbf{r}\right) +\frac{g}{2}\left| \psi \right| ^{2}-\Omega L_{z}\right] \psi d{\bf r}.
\end{equation}
Each stationary solution of the GPE of the form $\psi _{\mu}({\bf r},t)
=e^{-i\mu t}\phi({\bf r}),$ is a critical point of the energy subject to the
constraint of the norm, $\left. \frac{\partial E}{\partial \psi }\right|
_N[\psi_\mu]=0$.

\emph{Ground states.-} It is physically plausible to assume that the ENS
experiment actually produces condensates which are close to the \emph{absolute
  minimum of the energy} subject to some constraints. In practice, this means
that any theoretical effort to reproduce the experimental results must be based
on a method to find these \emph{ground states} of Eq.  (\ref{energy}) for given
norm and rotation speed. To achieve this goal and to be able to make
quantitative predictions we have worked numerically with the energy functional
[Eq.  (\ref{energy})], using a method known as Sobolev gradients. This method
is a preconditioned descent technique which allows to get accurate
approximations to the ground state where other minimization procedures fail. We
have applied this method over a Fourier basis with $32\times 32\times 64$
modes, which is enough to obtain accurate estimates of macroscopic values, such
as the angular momentum and the lowest critical frequencies. Nevertheless, the
most controversial results have been confirmed on a grid with $64\times 64\times
128$ modes.

\begin{figure}
\begin{center}
  \epsfig{file=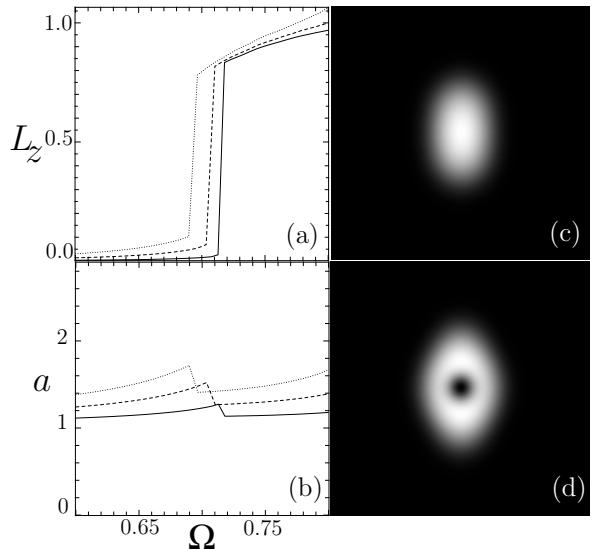,width=0.9\linewidth}
\end{center}
\caption{\label{fig-ens}
  (a-b) Angular momentum, $L_z$, and shape factor, $a=\langle
  y^2\rangle/\langle x^2\rangle$, of the GS of $E(\psi)$ [Eq. (\ref{energy})]
  as a function of $\Omega $. Solid, dashed and dotted lines correspond to
  $\varepsilon=0.03,0.06,0.09$ respectively. (c-d) Transverse density plots of
  a GS without vortex ($\Omega=0.65,\varepsilon=0.1$) and a GS with one vortex
  ($\Omega=0.75,\varepsilon=0.1$). Pictures are 12$\times$12 adimensional units
  large.}
\end{figure}

Our simulations are summarized in Figs. \ref{agujero} and \ref{fig-ens}. The
first striking prediction is that vortex lines are nucleated with a deformed shape,
like a tight string which is pulled from the middle as rotation is increased
[Fig. \ref{agujero}].  The second important result [Fig. \ref{fig-ens}(b),(d)]
is that even for a weak trap anisotropy, $\varepsilon=0.1$ [Fig.
\ref{fig-ens}(d)] a transverse section of the condensate shows a very deformed
cloud: the rotating condensate amplifies the small transverse deformation which
is used to transfer angular momentum to the condensate.

Furthermore, Fig. \ref{fig-ens} reveals that our numerical simulations confirm
several other features already found in the experiment. First, the critical
frequency for the nucleation of a vortex $\Omega_1$ is larger than what it 
is expected in the spherically symmetric trap. According to Fig. \ref{fig-ens}(a) in our
simulations (see parameters above) the condensate remains in a vortexless state
until $\Omega = 0.7(1)\times \omega _\perp/2\pi ,$ that is, $\Omega_1\simeq
162$ Hz, which is close to the value of 147 Hz obtained experimentally
\cite{ENS}, but far from the Thomas-Fermi estimate for radially symmetric
clouds.

The second remarkable feature is that, at $\Omega = \Omega_1$ the angular
momentum per particle $\langle L_z \rangle/N$ jumps to a value which is smaller
than 1 and which grows continuously until a second vortex is nucleated [Fig.
\ref{fig-ens}(a)].

Thus, the numerical simulations are in good agreement with the experience and
both contradict previous ideas: (i) the nucleation frequency is much larger than
expected in Thomas-Fermi theories, and (ii) the angular momentum has a
continuous, nonlinear dependence on $\Omega$. To obtain more insight on the
physical reasons for this behaviors we will analyze several simplified
situations.

\emph{Critical frequencies for axially symmetric traps .-} Let us first
consider the case of an axially symmetric trap. For these traps it is usually
thought that at a certain $\Omega = \Omega_1$ the system experiences a
transition from a vortexless state to a \emph{symmetric} state with one vortex.
Both states do also exist in the nonrotating case and have energies $E_0$ and
$E_1 > E_0$, respectively. This leads to the usual estimate of 
the critical frequency $\Omega_1 \simeq E_1-E_0$.

\begin{figure}
\begin{center}
  \epsfig{file=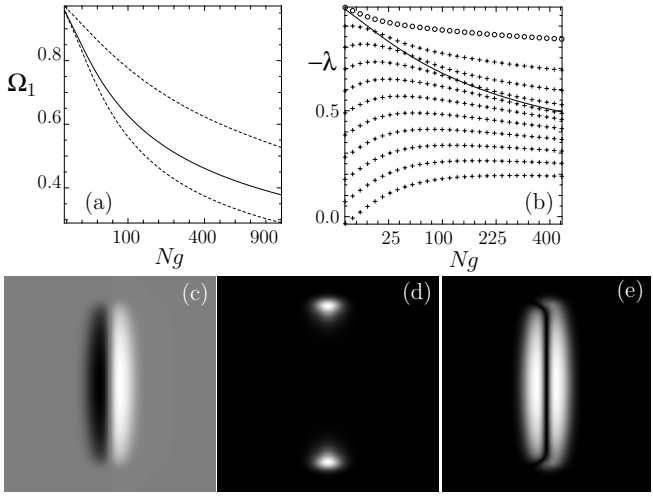,width=\linewidth}
\end{center}
\caption{\label{fig-omega}
  (a) Critical frequency, $\Omega_1$ for a spherical trap ($\gamma =
  \omega_\perp^2/\omega_z^2 = 1$, solid line), a cigar shape trap ($\gamma
  =19$, upper dashed line) and a pancake trap ($\gamma =1/19$, lower dashed
  line). (b) Critical frequency $\Omega_1$ (solid line) and eigenvalues arising
  from the linearization of the energy (circles) for a trap with $\gamma=19$.
  It is remarkable that there is always a negative eigenvalue which is not
  suppressed for $\Omega = \Omega_1$. (c) Vortex, (d) main destabilizing mode and (e)
  bended vortex line arising from a combination of both, for a trap with
  $\gamma=19$.}
\end{figure}

Taking the previous idea for granted and to learn how the {\em elongation of
  the trap} affects the nucleation of the first vortex, we have searched
the vortexless state and the state with one vortex in different {\em axially
symmetric} traps ($\varepsilon =0$), computing $\{E_0,E_1,\Omega_1\}$.  The
result of these calculations is shown in Fig. \ref{fig-omega}(a) for different
elongations.  There we see that the critical angular speed of a cigar shape
trap is larger than that of spherically symmetric and pancake traps.  The
explanation is simple: in an elongated trap, $\omega_z \ll \omega_\perp$, the
system expands preferably along the axial dimension in order to accommodate
more bosons and counteract the interaction energy. As $\omega_z$ is made
smaller this tendency becomes more evident, since the transverse shape no
longer satisfies the Thomas-Fermi approximation in which the second order
derivatives are negligible. Instead, the transverse shape of the cloud is
closer to a harmonic oscillator wavefunction, and as such the energy required
to introduce a single vortex, $\Omega_1$, becomes larger than in the
spherically symmetric case. In other words, the more elongated the trap is, the
more dilute the condensate seems on the XY plane, the closer $\Omega_1$ becomes
to its linear limit $\omega_\perp$, and the faster the condensate must spin to
accept a vortex. This effect, which can be characterized as a dynamical
separation of longitudinal and transverse degrees of freedom \cite{Perez-Garcia},
manifests in a departure of $\Omega_1$ from any Thomas-Fermi based results,
which correspond to a regime ruled by the transverse self-interaction of the
condensate.

\emph{Bended vortices in axially symmetric traps.-} An important feature of
elongated traps is that they require very little energy to produce longitudinal
excitations of the condensate \cite{us-vortex}. This makes feasible for small
perturbations to induce longitudinal modes and distortions of the vortex lines
\cite{Fetter-new,vortex-lines}. Thus our next step is to perform a linear 
stability analysis of the straight vortex line which we have already found.

We are looking for a quadratic expansion of the energy of a perturbed vortex.
Since the vortex $\psi_{vort}$ is a stationary state, this energy must read, up
to second order perturbations
\begin{equation}
\label{expansion}
E(\psi_{vort}+\epsilon\delta) \simeq E_1 - \Omega +
\epsilon^2 \sum_i \left(\lambda_{i,m}-m\Omega\right) |c_{i,m}|^2
\end{equation}
Here, $\lambda_{i,m}$ are the different eigenvalues which arise from
linearizing the GPE [Eq. (\ref{GPE-rot})] around $\psi_{vort}$, $m$ are their vorticities with
respect to $\psi_{vort}$ (See \cite{us-vortex}), and $c_{i,m}$ is the weight of
the $|i,m\rangle$ mode in the expansion of $\delta({\bf r})$. By finding the lowest
negative eigenvalue, $\lambda_{i,-1}<0$, we know that at a rotation speed
$\Omega_{meta}=|\lambda_{i,-1}|$ the energy (\ref{expansion}) becomes positive
for all perturbations and the vortex is locally stable or \emph{metastable}.

These calculations are summarized in Fig. \ref{fig-omega}(b). There we plot all
eigenvalues $-\lambda_{i,-1}$ and find that there is a number of them such that
$-\lambda > (E_1-E_0)$. This result suggests that there exists a range of
speeds $\Omega\in(\Omega_1=E_1-E_0,\Omega_m)$ in which the system nucleates a
vortex, but that vortex does not have a symmetric shape.  Instead, the
stability analysis points out that this vortex is bended by its ends, forming a
curved vortex line [Fig. \ref{fig-omega}(e)]. Although this is a perturbative
prediction arising from the linear stability analysis, these solutions are
consistent to what we formerly obtained by minimizing the full Hamiltonian and
which was explained above [Fig. \ref{agujero}].

The existence of a ground state with such a bended shape is one of the main
predictions of the paper, since it presents a symmetry
breaking due solely to rotation. In some sense the symmetry is broken twice: First because a 
phase asymmetric structure, the vortex, appears, and a second time because the vortex bends in some 
specific direction. We have
also observed that as the frequency is increased from $\Omega_1$ the vortex
bending becomes less and less evident (And $\langle L_z\rangle$ approaches
$N$).  However, when the threshold $\Omega_2$ for 
the nucleation of two vortices is reached the bending becomes again very important.

The existence of bending in the vortex lines may be checked in the experiment
and is probably related to the experimental observation that vortex lines appear to be
partially filled when viewed from above.

\emph{Role of transverse asymmetries.-} In Ref. \cite{ENS} the trap is
distorted over the transverse directions to induce some mechanical response of
the condensate, an effect which must be taken into account.

As it was already shown \cite{us-asym} any initial asymmetry of a condensate is
strongly emphasized by the rotation of the trap, an effect that may be
estimated analytically in the linear and Thomas-Fermi limits. These estimates
lead to a value of the angular momentum for a vortexless condensate and a
different one for a condensate with a single, centered, unit-charge vortex
line. The main difference with respect to the symmetric case is that a
vortexless state may acquire some angular momentum and that the nucleation of a
vortex line gives an increase of the total angular momentum which is inversely
proportional to the asymmetry factor, $\varepsilon$. From a practical point of
view this means that the system may ``decide'' to wait for a larger rotation
speed $\Omega_1>\Omega _{1,sym}$ before nucleating a vortex. Actually what
happens is that $\Omega_1/\mathrm{min}(\omega_x,\omega_y)$ increases with
respect to $\varepsilon$, but $\Omega_1/\omega_\perp$ decreases [Fig.
\ref{fig-ens}].

Although it is clear that the effect of transverse asymmetries is to provide a
larger critical frequency than expected it is not possible to use any analytic
estimate of the critical speeds because of the elongation of the trap and its
transverse deformation. Furthermore, due to the bending of the vortex lines it
is not possible to work with the corrections from \cite{us-asym} to get a
reliable estimate of $\Omega_1$.

\emph{Discussion and conclusions.-} We have used the GPE
to study theoretically the nucleation of vortices in an elongated trap similar
 to that of the ENS experiment \cite{ENS}. We have found several
mechanisms which lead to an increase of the angular speed which is required to
create those vortices and have found the remarkable feature that vortex lines
are bended by their ends. This provides some kind of symmetry breaking
bifurcation in the case of a single vortex \cite{yasabia}.  Recalling previous
experience with liquid helium, we suspect that this bending has some important
consequences, and that it might be responsible for the breakdown of
superfluidity in BEC by a mechanism which consists in several vortex lines
joining together and forming a turbulent pattern.

Using only the special geometry of the experiment we have obtained a value of
$\Omega_1=0.7(0)\omega_\perp$ which is close but above the experimental value.
However, this means that the experiment implies a systematic negative shift
both in $\Omega_1$ and in the angular speed at which the superfluid escapes,
$\Omega_c$.  In the experiment $\Omega_c \simeq 2\pi \times 210$ Hz, while the
theory predicts $\Omega_c = \omega_\perp (1-\varepsilon)$. If we trust the
value of $\Omega_c$ from the experiments $\Omega_c$ and make $\omega_{\perp} =
2\pi\times210$ Hz, our prediction is $\Omega_1=2\pi\times147$ Hz, which
coincides exactly which the experimental measures \cite{ENS}.

The fact that simple mean field theories are still valid when vortices enter
into the dynamics, and that we may still expect quantitative predictions based
on this simple models is interesting and has been shown in many other phenomena
\cite{PRL2000}. The striking result that the combined effect of asymmetry and
nonlinearity may lead to the phenomenology here studied is another example of the
richness of behaviors that one may expect from these nonlinear Quantum
mesoscopic systems.

This work has been partially supported by CICYT under grant PB96-0534.


\begin{thebibliography}{99}
\bibitem{Siempre}{M. H. Anderson, {\em et al.}, Science. {\bf 269}, 198 (1995);
    K. B. Davis, {\em et al.}, Phys. Rev. Lett. {\bf 75} (1995) 3969; C. C.
    Bradley, {\em et al.}, Phys. Rev. Lett. {\bf 75}, 1687 (1995).}
 
\bibitem{vortices_real}{P. G. Saffman,``Vortex dynamics", Cambridge University
    Press (1997); Y. Kivshar, B. Luther-Davis, Phys. Rep.  \textbf{298}, 81
    (1998).}
  
\bibitem{vortices-BEC}{D. S. Rokhsar, Phys. Rev. Lett. {\bf 79} (1997) 2164; R.
    J. Dodd {\em et al}, Phys. Rev. A {\bf 56} (1997) 587; T. Isoshima and
    K. Machida, Jour. Phys.  Soc. Jpn. {\bf 66} (1997) 3502; R. Dum, {\em et
      al.}, Phys. Rev. Lett. {\bf 80} (1998) 2973; A. Svidzinsky and A. L.
    Fetter, Phys. Rev.  A {\bf 58} (1998) 3168; F.  Zambelli, S. Stringari,
    Phys. Rev. Lett. {\bf 81} (1998) 1754; A. L.  Fetter, J. Low Temp. Phys.
    {\bf 113}, 189 (1998); E. L. Bolda, D. Walls, Phys.  Rev. Lett. {\bf 81} (1998)
    5477; H. Pu {\em et al.}, {\bf 59} (1999) 1533; M.  Caradoc-Davies, R. J.
    Ballagh, and K. Burnett, Phys. Rev. Lett {\bf 83}, 895 (1999); Phys. Rev. A
    {\bf 60} (1999); T. Isoshima and K.  Machida, Phys.  Rev. A {\bf 59} (1999)
    2203; Phys. Rev. A {\bf 60}, 3313 (1999); D. L. Feder, C. W. Clark and
    B. I.  Schneider, Phys. Rev. Lett. {\bf 82} 4956 (1999); M. R. Matthews
    {\em et al.}, Phys. Rev. Lett. {\bf 83} 2498 (1999); D. L. Feder, C. W.
    Clark and B. I. Schneider, Phys. Rev. A {\bf 61} 011601 (2000).}
  
\bibitem{Rokshar}{D. A. Butts and D. S. Rokhsar, Nature {\bf 397}, 327
    (1999).}
  
\bibitem{us-vortex}{J. J. Garc\'{\i}a-Ripoll and V. M. P\'erez-Garc\'{\i}a,
    Phys. Rev. A {\bf 60}, 4864 (1999).}
  
\bibitem{PRL2000}{J. J. Garc\'{\i}a-Ripoll, V. M. P\'erez-Garc\'{\i}a, Phys.
    Rev. Lett. {\bf 84} (2000) 4264; V. M. P\'erez-Garc\'{\i}a, J. J.
    Garc\'{\i}a-Ripoll, Phys. Rev. A (to appear) (2000).}
  
\bibitem{ENS}{K. W. Madison, F. Chevy, W. Wohlleben, J. Dalibard, Phys. Rev.
    Lett. {\bf 84} 806 (2000).}
  
\bibitem{Lundh}{E. Lundh, C.J. Pethick, H. Smith, Phys. Rev. A \textbf{55},
    2128 (1997).}
  
\bibitem{Perez-Garcia}{V. M. P\'erez-Garc\'{\i}a, H. Michinel, H. Herrero,
    Phys. Rev. A {\bf 57} (1998) 3837; D. Margetis, J. Math. Phys. 40, 5522
    (1999); O. Zobay {\em et al.}, Phys. Rev. A 59, 643 (1999); E. A.
    Ostrovskaya {\em et al.}, Phys. Rev. A 61, 031601(R) (2000); L. D. Carr
    {\em et al.}, {\tt cond-mat/0004287}.}
  
\bibitem{Fetter-new}{A. S. Svidzinsky, A. Fetter, preprint.}
  
\bibitem{us-asym}{J. J. Garc\'{\i}a-Ripoll and V. M. P\'erez-Garc\'{\i}a,
    {\tt cond-mat/003451}.}
  
\bibitem{vortex-lines}{A systematic theory for the bending of vortex lines has been developed for the 
Ginzburg-Landau equation in: J. P. Keener, J. J. Tyson, SIAM Rev. 34 (1992) 1; M.
    Gabbay, E. Ott, P. Guzdar, Physica D {\bf 118} (1998) 371.}
  
\bibitem{yasabia}{When there is more than one vortex the symmetry breaking in
    the form of an array of vortices has been known for some time. See e.g.
    Refs. \onlinecite{Rokshar,us-asym}.}

\end{thebibliography}
\end{document}